# Experimental Demonstration of Nanosecond-Accuracy Wireless Network Synchronization

Marcelo Segura, S. Niranjayan, Hossein Hashemi, *Senior Member, IEEE* and Andreas F. Molisch, *Fellow, IEEE*

*Abstract*— Accurate wireless timing synchronization has been an extremely important topic in wireless sensor networks, required in applications ranging from distributed beam forming to precision localization and navigation. However, it is very challenging to realize, in particular when the required accuracy should be better than the runtime between the nodes. This work presents, to our knowledge for the first time, an experimental timing synchronization scheme that achieves a timing accuracy better than 5-ns rms in a network with 4 nodes. The experimental hardware is built from commercially available components and based on software-defined ultra-wideband transceivers. The protocol for establishing the synchronization is based on our recently developed "blink" protocol that can scale from the small network demonstrated here to larger networks of hundreds or thousands of nodes.

## I. Introduction

Time synchronization in distributed wireless networks has been studied for the past two decades, as it is an essential component of many applications. The required accuracy, and the resulting algorithms, differ according to the application. In many cases, such as multiple-access protocols, timing synchronization within less than one packet duration (on the order of a millisecond) is sufficient. In other applications such as diversity transmission, timing synchronization within one symbol duration (on the order of microseconds) is desired. For these applications, the IEEE 1588 standard is widely accepted as a solution [1]. Comparable algorithms that have been explored include Reference Broadcast Synchronization (RBS) [2], Timing Sync Protocol for Sensor Networks (TPSN) [3] and Flooding Time Synchronization Protocols (FTSP) [4]. In general, these timing synchronization algorithms are based on packet exchange, where the accuracy is related to the symbol or preamble length and the propagation delay between the nodes is neglected [5].

This work was supported in part by the Office of Naval Research Electronic Warfare Science and Technology program, and the Ming Hsieh Institute at the University of Southern California.
M. Segura, H. Hossein and A. F. Molisch are with the Department of Electrical Engineering, University of Southern California, Los Angeles, CA 90089, USA. e-mail: ({mjsegura ; hosseinh ; molisch}@ usc.edu).
S. Niranjayan, was with University of Southern California, Los Angeles, CA 90089, USA, now with Amazon (e-mail: sniranjayan@gmail.com).

There are, however, also a variety of applications that require much better timing synchronization accuracy, in particular better than the runtime of the signal between the nodes. Such applications, such as coordinated jamming, distributed beam forming, fine grain localization, tracking, and navigation, require timing synchronization with nanosecond precision or even better. In the absence of differential GPS (which is often the case in many military and civilian scenarios), maintaining such accurate timing synchronization in large wireless networks is challenging; because, the timer in each node is derived from an independent oscillator that is affected by random drifts and jitter [6]. To maintain nanosecond-level timing synchronization in a network (1) timing deviation between pairs of nodes must be measured accurately with nanosecond precision, and (2) fast correction algorithms must be applied across the entire network [9, 16]. This paper offers solutions and experimental demonstrations for both of these components.

Ultra-Wide Band (UWB) signals are used to precisely extract the timing information between the nodes due to their accurate distance measurement capability as well as superior resiliency to multi-path effects [7]. UWB systems have been used previously to demonstrate high-precision timing synchronization between two nodes. In [8], a commercial IEEE 802.14.5 radio was employed to obtain an accurate Time of Arrival (TOA) detection algorithm; the authors also proposed an Adder-Based Clock (ABC) approach for timing synchronization and implemented a prototype to show the accuracy of the synchronization block with nanoseconds precision. However, no network experiments were carried out. As a matter of fact, to the best of our knowledge, there are no previous prototypes that demonstrate network synchronization with nanosecond accuracy.

The challenge in achieving timing synchronization in a wireless network lies in the required high speed of the node-to-node timing inaccurate measurements and fast corrections across the entire network. In this work, a recently developed algorithm [9, 16], called "*Blink*" that enables network timing synchronization without requiring an external broadcast signal from a coordinator is implemented. The *blink* algorithm uses a consensus approach such that timing information propagates through the network, while timing errors are averaged, and exploits the path diversity present in the network. Simulations demonstrate excellent scalability, such that the obtained timing precision in large (hundreds of nodes) networks is only marginally worse than those in two-node setups [9].

The current paper demonstrates and validates through a set of experiments the very accurate network synchronization that was predicted in simulations and shows that it is possible to achieve this on a prototype using commercial components, *i.e.*, without requiring custom Integrated Circuits (IC). The main components of this work are (1) a software defined ultra wideband transceiver node implemented with Commercial-Off-The-Shelf (COTS) components, capable of high accurate synchronization, (2) a fast re-timing algorithm implementation on a Field Programmable Gate Array (FPGA), and (3) a test bed network consisting of three slave nodes and one master that facilitates the validation of this algorithm and future improved versions.

The remainder of the paper is organized as follows. Section II presents a formulation for network timing synchronization. Section III presents summary of the previously-published [9] *blink* algorithm. Sections IV and V describe the implemented node prototype and the hardware implementation on FPGA of the *blink* algorithm, respectively. Section VI presents the simulation and experimental results that prove the stability and accuracy of the network timing synchronization. Finally, some conclusions are presented.

## II. PROBLEM OVERVIEW

Consider a large network with $N_s$ slaves nodes with inaccurate internal oscillators, and $N_m$ master nodes with accurate internal oscillators, which will serve as timing references and initiators in the distributed algorithm. Let $C_k(t^*)$ denote the current local timer value of a particular node $k$ at a given absolute time instance $t^*$. The network timing problem consists of maintaining, at all times, the offset between nodes below a predefined threshold $\sigma_{tol}^2$ as

$$E\{(C_k(t^*) - C_j(t^*))^2\} < \sigma_{tol}^2, \forall i \neq j. \quad (1)$$

A major challenge lies in the fact that due to the random deployment, the network topology is unknown. Furthermore, in order to be scalable and robust to changes, the solution should be distributed. Finally, in a large network, broadcast transmission of timing information from a master to all other nodes in the network is not possible. Therefore, the timing information propagates through the network aggregating errors.

## III. BLINK ALGORITHM DESCRIPTION

A brief review of the *blink* algorithm is presented in this section. More details and comparison with other distributed timing synchronization schemes are given in [9] and implementation is discussed in [16].

Inside the physical network, a virtual network called *Timing Virtual Network* (TVN) is defined. The TVN consist of nodes, links and the fast re-sync algorithm that maintains the timing in the network. Different physical layers may be used for timing and communications. The current work focuses on the timing physical layer; the implementation of the communication layer with a certain packet structure (*e.g.*, using commercial IEEE 802.15.4 transceivers) is beyond the scope of this paper, but has been considered in the *blink* protocol definition. The topology of the network is defined by the link matrix $L$ as follows:

$$\{L_{i,j}\} = \begin{cases} 0, & \text{if } i = j \text{ or } \gamma_{ij} < \gamma_{th} \\ 1, & \text{if } \gamma_{ij} \geq \gamma_{th} \end{cases}, \quad (2)$$

where $\gamma_{ij}$ defines the Signal to Noise Ratio (SNR) between nodes and $\gamma_{th}$ is the minimum SNR needed to establish a connection. Each node is assigned to a tier $T_i$ (Fig. 1.) and by default masters are in tier 0. The blink algorithm consists of two phases. In Phase I, the TVN is created and initialized and in Phase II, the network timing is maintained through continuous consensus-based corrections.

### A. Phase I

All nodes (slaves and masters) learn the propagation delays (pseudo-ranges) from their neighbors and record the values. In a fresh deployment scenario, the network topology and the tier structure is unknown. Therefore, propagation delays are acquired in a distributed fashion, using a modified version of Carrier Sense Multiple Access Collision Avoidance (CSMA/CA). In this phase, the slave nodes also get assigned to different tiers. Tier 1 consist of nodes that can derive timing directly from the master; the size of this tier can be limited by the number of slave nodes that have an acceptable SNR to the master or the number of slave nodes that can communicate with the master with acceptable time constraints (this is important for scalability of the system). Other tiers are also defined consecutively in a similar fashion (Fig. 1).

In this paper, the main objective is to experimentally demonstrate the stability and accuracy of the network timing synchronization. Therefore, it is assumed that nodes in tier 1 have already acquired timing from the master, and the focus is on demonstration of slave-slave timing propagation and synchronization.

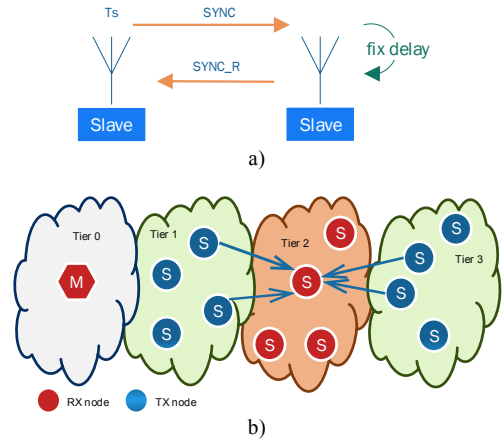

Fig. 1. a) Simplified node-node time exchange scheme; b) TVN topology after Phase I.

### B. Phase II

The master initializes the *blink* cycles transmitting a timing signal; in this implementation, a length 31 m-sequence is used for the timing signal. Slave nodes receive this sequence, perform correlation and peak detection to extract the timing information, use the result to correct their internal timers, and transmit the same sequence according to their tier associations. The *blink* algorithm proposes that nodes on odd (or even) tiers transmit simultaneously. In the current hardware implementation, in order to guarantee orthogonality of the nodes' signals, a TDMA approach was implemented. In the case of large networks, and to preserve the distributed characteristics of the algorithm, random slots can be assigned to different tiers. In this work, each node transmits on the slot time associated with its tier. If the number of adjacent nodes with established connection between tiers is denoted by $N_n$ and the slot time is $\tau_{slot}$, then the blinking cycle will take a total time of $(N_n - 1)\tau_{slot}$.

During the blinking cycles the slave nodes, after receiving the timing signals from their neighbors, measure the TOA relative to their own local timers, $\{t_k\}_{k=1}^{Nn}$, where $t_k$ is the measured TOA of node $k$. Using the learned propagations delays from phase I, $\{\tilde{t}_k\}_{k=1}^{Nn}$, each node compares the two values and compute its own offset related to all its neighbors as $\{t_k - \tilde{t}_k\}_{k=1}^{Nn}$. The nodes then use a consensus algorithm where each neighbor node can be weighted with a different factor to provide a timing correction value as

$$\mathcal{T}_o = \sum_{k=1}^{Nn} w_k (t_k - \tilde{t}_k), \qquad (3)$$

where $w_k$, represent the weighting factor for each link. The blinking cycles continue indefinitely. Reference signals from the master "pull" the timing in the network to agree with the master timing. Timers on tier one nodes are influenced by master reference signals as well as neighbors and the timing information from there propagates through the network.

Figure 1.b depicts the working of the algorithm. At the beginning, in TDMA slot 1, nodes on Tier 1 are in transmitting mode and nodes on Tier 2 and Tier 3 are on receiving mode. Due to $\gamma_{th}$ restrictions, the nodes on Tier 3 do not detect the signals from Tier1 and therefore do not correct their timers. During the next timeslot, the nodes on Tier 2 transmit and the others nodes listen. The nodes on Tier 2, after receiving all the diversity information from their neighbors, correct their internal timers based on the timing correction value calculated from (3). This process is repeated every blinking cycle on each tier accordingly with the TDMA slots.

## IV. Prototype Node Hardware

As a proof of concept, to demonstrate the algorithm capabilities in real time, custom nodes using UWB as physical layer signals are implemented. UWB signals are selected for timing purpose due to the inverse relation between the TOA estimation error and signal bandwidth. The hardware nodes are software programmable to provide flexibility and rapid prototyping given that the timing synchronization algorithm is implemented completely on digital hardware. The downside of this approach is that timing accuracy will be limited to the performance of available commercial hardware (to be discussed later). Future custom realization of the node hardware enables more accurate network timing synchronization.

Each node is composed of two principal elements: a high speed Analog to Digital Converter (ADC) board from Texas Instruments and a Xilinx Kintex Field Programmable Gate Array (FPGA) evaluation board (Fig. 2).

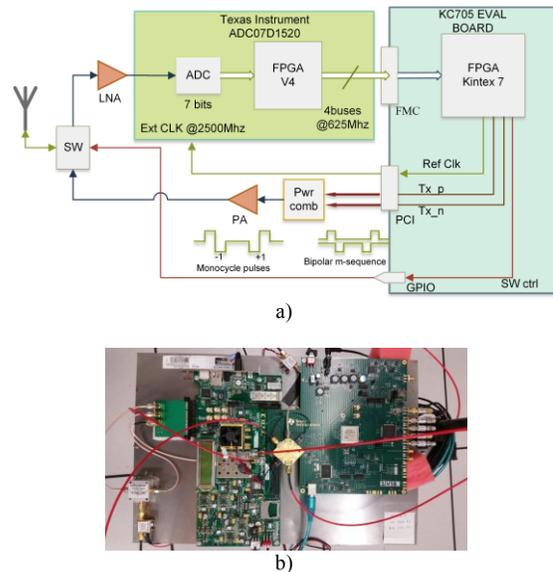

Fig. 2. The (a) architecture block diagram and (b) photo of the prototype hardware for each node.

The receiver is implemented using the ADC07D1520 board that can run at a maximum sampling rate of 3 GSps with 7 bits of resolution. The available FPGA Virtex 4 in this ADC board is used as a simple pass-through buffer (no computation). In order to reduce the frequency of the digital signals, the ADC output provides four interleaved buses running at a quarter of the sampling frequency. In the current implementation, due to internal FPGA clock manager restrictions, the sampling frequency is set to 2.5 GSps. This reference clock is provided by the Kintex board to the ADC thought a PCI-SMA connection. The ADC samples the incoming signals at RF, after amplification with a Low Noise Amplifier (LNA), without any frequency down-conversion. While flexible and consistent with the vision of a true Software Defined Radio (SDR), the limited speed and resolution of available commercial ADCs dictates the achievable timing accuracy.

The transmitter is implemented digitally using the Gigabit Transceivers (GTX) embedded in the FPGA without using frequency up-converters. A simple Binary Phase Shift Keying (BPSK) modulation is implemented using two GTXs with opposite polarity that will feed a power combiner. The GTX provide great flexibility to design monocycle signals changing the pulse width accordingly with the clock reference. In order to meet the Nyquist theorem and relaxing the sampling rate, the implemented monocycle pulses have a 1.2 ns width, providing approximately 800 MHz signal bandwidth. After the power combiner, the BPSK signal passes through a Power Amplifier (PA) and connects to a discone antenna [10], through an RF switch.

The developed node is clocked from a low-cost 25 MHz crystal oscillator with 50 ppm (part per million) accuracy that is located on the KC705 FPGA board. All the clocks that are internally used in the FPGA and the reference clock that feeds the ADC are created from the same crystal reference. The implemented UWB node, shown on Fig. 2, is fully digital, modular and easy to modify since all transmitter/receiver blocks are controlled by the Kintex FPGA.

## V. FPGA IMPLEMENTATION OF THE BLINK ALGORITHM

The accuracy of the *blink* algorithm is intrinsically related with the TOA estimation error. A well-known and simple method to estimate the signal TOA consists of peak detection after cross correlation [11]. It is also known that peak detection will not provide the right time under non-Line-Of-Sight (LOS) conditions; but, can be used as a starting point for further processing [12, 13]; however this is not implemented in our hardware yet. The blink algorithm proposes that nodes learn the channel multipath signatures, and use that information to create proper reference templates. In this implementation, in order to simplify the algorithm, the same template for all nodes in the network is used. Better results can be obtained if the nodes continuously update their reference templates as a function of propagation channels variations.

The timing signal used for the TOA estimation must provide a high ratio for the autocorrelation peak to the side-lobe peak. An m-sequence of length 31 was selected for the timing signals since it meets the minimum requirements. The length of the sequence is constrained by the correlator size feasible with the existing hardware. Longer sequences give higher peaks; but, increase the complexity of logic hardware. The sequence transmission is controlled by the TX Logic block as shown in Fig. 3. As previously explained, a TDMA scheme was implemented where each node transmit on different slots according to the node identification and tier association to achieve better peak detection and reduce the interference between nodes on the same tier.

The most logic-consuming and timing-constrained algorithm to be implemented in the FPGA is the parallel correlator, because it has to run in real time and its size increases with the sequence length and the ADC resolution. Hence, to meet the timing constraints imposed by the FPGA logic, a dual data rate input is implemented, which transforms the 4 buses coming from the ADC at 625 MHz into 8 buses running at 312.5 MHz. Furthermore, to reduce complexity and satisfy real time operation, 4 bits (sign plus 3 bits magnitude) are taken from each bus. The correlator was implemented in parallel following an architecture called PTT (Parallel samples, Parallel Coefficients, Time division multiplexing). This architecture is highly efficient for an FPGA implementation and works as follows. On each clock cycle, a $s*k$ array, where $s$ is the number of samples in parallel and $k$ is the number of correlation points calculated simultaneously, is processed by the PPT correlator. On the next clock, another array is processed and after $L/s$ cycles, where $L$ is the template length, the correlation is completed [14]. In the current implementation, considering the sampling rate of 2.5 GHz and the sequence length of 62 ns, the template length will be $L=160$; therefore, the correlator has 20 arrays as those described in Fig. 4. In the implemented design, $s=8$, $k=8$, and each sample and coefficient has a 4-bit width; therefore, in each clock cycle, 64 multiplication have to run in parallel for every array. The major challenge in this design is satisfying the zero latency restriction for samples propagation (vertical arrows in Fig. 4) between multipliers in the same array given the high sampling rate. Finally, the outputs of the parallel correlator feed the peak detector block.

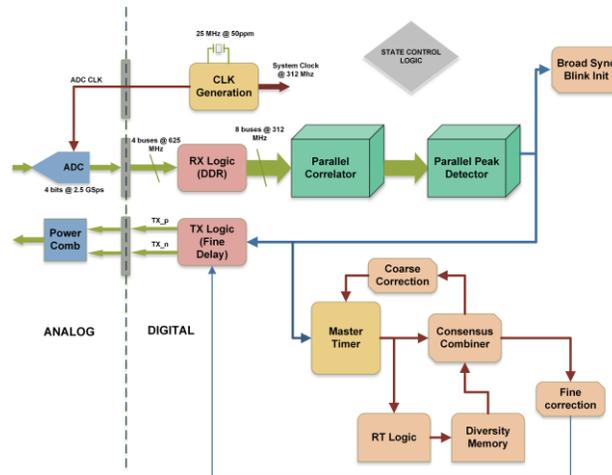

Fig. 3. Blink algorithm block diagram. The dash line shows the boundary between the analog hardware and the digital algorithm implementation inside the FPGA.

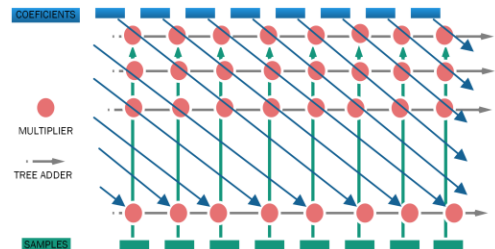

Fig.4. Parallel correlator array.

The peak detector compares each branch at the correlator output against the $\gamma_{ij}$ threshold. In the current proof-of-concept experiments, static or quasi-stationary channel conditions are assumed; but, this detector can be easily updated to handle channel variation implementing a dynamic threshold algorithm like that one proposed in [15]. The peak detector determines the time and corresponding branch at which a correlation peak happens. This information offers two resolution levels corresponding to (1) the coarse correction, that will be applied to the master timer and (2) the fine correction that will be used to adjust the transmission time. The resolution of coarse correction is limited by the system clock at which the correlator is running, in our case 312 MHz or 3.2ns. The resolution of the fine correction is limited by the rate of the GTX that is 0.2 ns in this design. The current experiments were conducted without controlling the fine error; this will be included in further implementation.

During the learning process in Phase I, the nodes estimate the propagation delay, $\{\tilde{t}_k\}_{k=1}^{Nn}$, between their neighbors. This phase of the algorithm is implemented by averaging the roundtrip measurements between the nodes. The roundtrip measurement works as follow: (1) node A resets its timer and transmit a sequence, (2) node B uses the correlator and peak detectors to determine the TOA and immediately resets its timer and triggers the TX Logic, (3) node A captures its timer value after peak detection. The process is repeated 16 times to reduce errors in the TOA estimation and the average value is stored at the "Diversity Mem" block. The hardware implementation of the roundtrip procedure is shown in Fig. 3 as "RT Logic".

During the blinking cycles, when a peak is detected, the master timer captures its current value that represent the estimated $t_k$. Depending on their tier location, the node waits for the timing diversity update from its neighbors, as shown on Fig. 1, and then (3) is computed using the previously learned values $\tilde{t}_k$, that are stored on the diversity memory block. As a result, the offset coarse error, $\mathcal{T}_o$, is obtained and the master timer is corrected. Lastly, the fine error will be applied to the "TX Logic" as a shift in the sequence to be transmitted (not implemented yet). The node will transmit again, once the offset correction is done and following the TDMA scheme.

Finally, the "Broad Sync Logic" block implements the initialization procedure of blinking cycles described in Section III B, and the "Control State Machine" block controls all the phases involved in the *blink* algorithm.

## VI. SIMULATION AND EXPERIMENT RESULTS

The *blink* algorithm was fully implemented using the Matlab System Generator, a tool that allows running real-time hardware simulations in a Simulink environment. Simulations indicate the anticipated performance prior to actual hardware experiments. Current setup consists of three slave nodes and one master node. In order to consider the channel multipath signature, a received signal was recorded and used in the simulations. The distance between the nodes is simulated by the channel propagation delays. In the presented simulations, Node 0 belongs to tier 1, Node 2 belongs to tier 2 and Node 1 belongs to tier 3. Simulations show that synchronization convergence time is very fast (25.6 $\mu s$). This time is measured from the beginning of Phase II of the *blink* algorithm until the time that offset error between nodes reach the best system resolution, which in our implementation is one period of the master timer clock. The fast convergence is reasonable as the described network is simple with only one node per tier and one master node.

Actual hardware experiments were conducted in three different environments, indoor office with LOS and NLOS, and outdoor environments. For the NLOS experiment, node 0 has direct LOS link with node 2 while the direct LOS link between nodes 1 and 2 is obstructed by a metal shelf (Fig. 6). In order to compare the simulations results against experiment, the network topology was the same as described in the simulation setup and the distance between the nodes are expressed in Table 1.

TABLE I
DISTANCE BETWEEN NODES IN METERS

|  | Master-N0 | N0-N2 | N2-N1 |
|---|---|---|---|
| Indoor (LOS) | 1 | 1.8 | 2.4 |
| Indoor (NLOS) | 0.9 | 1.6 | 1.7 |
| Outdoor | 1.9 | 3.7 | 4.5 |
| Simulation | 3 | 5.4 | 7.5 |

In order to measure the relative time offset between the synchronized nodes, an additional timeslot was introduced in our TDMA scheme where all the nodes transmit simultaneously accordingly with their internal timer. On this additional slot, each node transmits the timing sequence and a short pulse that will be used for measurement. If the slave nodes are synchronized, all these pulses should be aligned. This extra slot does not affect the *blink* algorithm since is can be consider as an additional tier in the network. The three pulses coming from the slave nodes are measured by connecting the nodes through identical coaxial cables to a high-speed four-channel real-time oscilloscope, Tektronix DPO71254B (Fig. 5.b). The oscilloscope is equipped with a statistical analysis tool that allows measuring the mean (offset) and standard deviation (jitter) of the signals received at its different channels. Node 2 is considered as reference to trigger the scope since it is the node that has the most timing diversity. Therefore, the statistics measured are carried out considering node 2 as the time reference. In Fig.5.a, the pulses captured buy the scope are shown with their time offsets.

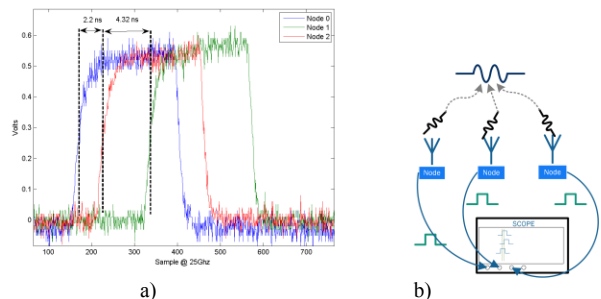

Fig. 5. a) Measured timing offset of the synchronized network.
b) Experimental setup.

The results on Table 2, show the mean values and standard deviation over four thousand captured pulses confirming that timing jitter values agree with the simulation predictions and equal to approximately one clock period (3.2 ns).

TABLE II
STATISTICAL ANALYSIS

|  | Indoor(LOS) | | Indoor(NLOS) | | Outdoor | |
|---|---|---|---|---|---|---|
|  | Node 0 | Node 1 | Node 0 | Node 1 | Node 0 | Node 1 |
| Mean (ns) | 0.118 | 0.886 | 0.820 | 2.812 | 1.554 | 2.822 |
| Std. dev. (ns) | 3.316 | 3.358 | 3.412 | 3.211 | 3.011 | 3.002 |

Evaluating the mean value results, indoor LOS experiment has the best performance due to high SNR, achieving sub-nanosecond offset accuracy. In the case of indoor NLOS experiment, the mean error of node 0 is better than node 1, since the NLOS condition reduces the TOA estimation accuracy.

For outdoor experiments, the timing offset increases due to degradation in TOA estimation directly related to lower SNR.

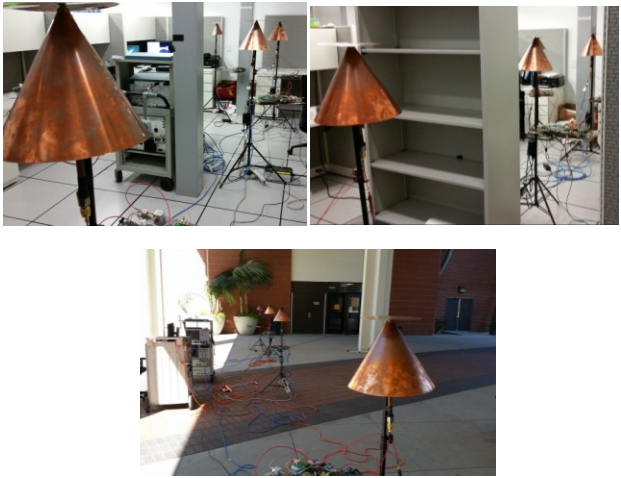

Fig. 6. Experiments setup: Top Left, indoor with LOS. Top Right, indoor with NLOS. Bottom, outdoor with LOS.

The standard deviation in this case, representing the timing synchronization jitter, is in line with the implemented coarse timing accuracy of the implemented system clock (3.2 ns). Therefore, it is expected that implementing fine timing correction and increasing the master timer clock will push the jitter under the sub-nanosecond range.

Finally, the self-synchronized wireless sensor network is setup to provide coherent waveforms for distributed beam-forming applications. An additional receiving antenna was positioned at an equal distance from all nodes and connected to the scope. Coherent addition of signals wirelessly received from different nodes (Fig. 7) is a nice demonstration of the benefits of the nanosecond network timing synchronization scheme.

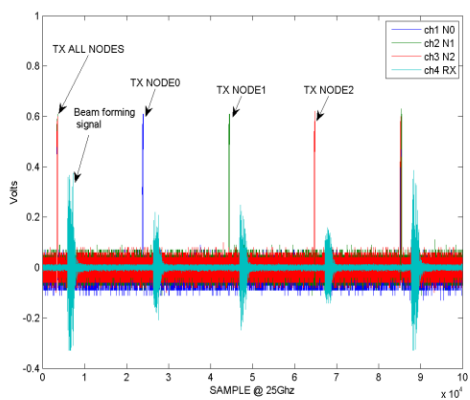

Fig. 7. Measured beam forming experimental results.

## VII. CONCLUSION AND FUTURE WORK

This work demonstrates experimentally that wireless network nodes can all synchronize in a cooperative and distributed manner with a timing accuracy that is markedly lower than the propagation delay between them. Under good SNR conditions, the network can attain sub-nanosecond accuracy as was demonstrated in the indoor experiments. The timing jitter in the synchronized network is limited primarily by the coarse correction applied in the master timer. The jitter can be reduced in future work by adopting optimal correlation architectures, increasing the frequency of the master timer and applying post signal processing. In the current implementation, the timing offset in the synchronized network is constrained by the resolution of the TOA estimation algorithm. In the future, offline signal processing to improve the detection capabilities and improve the synchronization under non-LOS conditions can be implemented. Finally, future hardware realizations can benefit from monolithic low-cost low-energy realization of the sensor nodes.